\documentstyle[aps,epsf,epsfig,preprint]{revtex}
\def\prl#1#2#3{{\it Phys. Rev. Lett.} {\bf #1}, #2 }
\def\pla#1#2#3{{\it Phys. Lett. A} {\bf #1}, #2 }
\def\pre#1#2#3{{\it Phys. Rev. E }{\bf #1}, #2 }
\def\pra#1#2#3{{\it Phys. Rev. A }{\bf #1}, #2 }
\def\prb#1#2#3{{\it Phys. Rev. B }{\bf #1}, #2 }
\def\cha#1#2#3{{\it Chaos }{\bf #1}, #2 }
\def\lanl#1{\it LANL archives }
\def\progth#1#2#3{{\it Prog. Theor. Phys. }{\bf #1}, #2 }
\def\prophy#1#2#3{{\it Proc. Phys. Soc. London A }{\bf #1}, #2 }
\def\jnsc#1#2#3{{\it J. Nonlinear Sci. }{\bf #1}, #2 }
\def\rmod#1#2#3{{\it Rev. Mod. Phys. }{\bf #1}, #2 }
\def\fund#1#2#3{{\it Fund. Math. }{\bf #1}, #2 }
\def\funct#1#2#3{{\it Funct. Anal.  Appl. }{\bf #1}, #2 }
\def\iter#1#2#3{{\it Int. J. Bifurcation and Chaos }{\bf #1}, #2 }

\def\jsp#1#2#3{{\it J. Stat. Phys.}  {\bf #1}, #2 }
\def\physd#1#2#3{{\it Physica D} {\bf #1}, #2 }
\def\snsa#1#2#3{{\it Proc. Ind. Natl. Sci. Acad.}{\bf #1} #2 }

\def\eg{e.g. }
\def\etl{et~al.}

\def\beqr{\begin{eqnarray}}
\def\eqnr{\end{eqnarray}}
\def\beq{\begin{equation}}
\def\bc{\begin{center}}
\def\ec{\end{center}}
\def\eqn{\end{equation}}
\begin{document}
\title{A Plethora of Strange Nonchaotic Attractors}
\author{Surendra Singh Negi and Ramakrishna Ramaswamy}
\address{School of Physical Sciences, Jawaharlal Nehru University, New Delhi 110 067, INDIA}
\date{\today}
\maketitle
\begin{abstract}
We show that it is possible to devise a large class of
skew--product dynamical systems which have strange nonchaotic
attractors (SNAs): the dynamics is asymptotically on fractal attractors
and the largest Lyapunov exponent is nonpositive. Furthermore,
we show that quasiperiodic forcing, which has been a hallmark of essentially all
hitherto known examples of such dynamics is {\it not} necessary for the creation of SNAs.
\end{abstract}

\newpage
\section{Introduction}

Since 1984, when Grebogi, Ott, Pelikan, and Yorke \cite{gopy}
described attractors that were strange but not chaotic, interest
in these exotic objects has been increasing \cite{rmp}. Several studies, both
theoretical \cite{theory,theory1,theory2,theory3,theory4,yl,yl1} and
experimental \cite{exp,exp1,exp2,exp3,newexp,newexp1},
have, over the years, elucidated the principal features
of such attractors which appear to be {\it generic} in
quasiperiodically driven nonlinear dynamical systems.
These are geometrically strange sets (fractals) on which {\it all}
Lyapunov exponents are either zero or negative. Further,
owing to their fractal character, motion on SNAs is intermittent and
aperiodic.

A very important connection between classically driven systems and
the Schr\"odinger equation for a particle in a quasiperiodic potential
\cite{bondeson} makes the correspondence between nonchaotic attractors
and localized states \cite{ks,prss,nr1}, and this further underscores the
interest in strange nonchaotic attractors (SNAs). These results show
that the transition from an invariant circle to SNA
in the iterative mapping is related to the transition from
extended to localized states in the quantum system.

Although there are many known examples of systems with SNAs,
proving the existence of such attractors is a mathematically
nontrivial task. Indeed, rigorous results exist for only two
systems---the original system introduced by Grebogi \etl, and the
Harper map \cite{ks}. Keller \cite{keller} and Bezhaeva and
Oseledets \cite{bo} have shown in the dynamical system \cite{gopy}
\begin{eqnarray}
\label{tanh}
x_{i+1} &=& 2 \alpha \cos 2\pi \phi_i \tanh x_i\\
\phi_{i+1} &=& \{\omega+\phi_i\},
\label{irra}
\end{eqnarray}
(we use the notation $\{y\} \equiv y$ mod 1) that
for $\alpha$ sufficiently large and for $\omega$
an irrational number,
the attractor is fractal, has a singular--continuous spectrum,
and is the support of an ergodic SRB measure \cite{ruelle}.
For the Harper map ($E$ being a parameter)
\begin{eqnarray}
\label{harpc}
x_{i+1} &=& -[x_i - E + 2\alpha \cos 2\pi \phi_i]^{-1} \\
\phi_{i+1} &=& \{\omega+\phi_i\},
\end{eqnarray}
a persuasive argument \cite{ks} suggests that,
again for $\alpha$ sufficiently large and $\omega$ irrational,
the attractor of the dynamics is fractal, and the dynamics
is nonchaotic since the Lyapunov exponent is nonpositive \cite{prss}.

Virtually all known examples of systems with SNAs---and the
two cases above are typical---have the skew--product
\footnote{In a skew--product system, one of the variables evolves 
entirely independently of the other(s), as $\phi$ does in these
examples.} form and have a quasiperiodic driving term.  How necessary are
these features?  This question has considerable practical relevance
since there are known cases of {\it experimental} systems
\cite{newexp,newexp1} where the dynamics appears to be on strange
nonchaotic attractors, but where there is no
quasiperiodic driving.  Furthermore, it is also known that
a judicious choice of forcing term can convert chaotic
dynamics to nonchaotic dynamics \cite{raj,sw,sw1,prinsa}.
Are the resulting attractors SNAs?
In particular, can SNAs be formed via non-periodic (and also not
quasiperiodic) driving of a nonlinear system?

In the present paper we address two issues that are of concern
in the study of SNAs. The first deals with the ubiquity of such
attractors: what are the necessary conditions for a dynamical
system such that SNAs result? While we do not provide an
exhaustive answer, our study shows that strange nonchaotic
dynamics can be realized in a wide variety of systems, and
may in fact be quite common. Our examples all remain,
however, within the skew--product paradigm. The second issue relates, as
discussed above, to the necessity of quasiperiodic forcing and
we find that it is possible to construct dynamical systems
wherein there is {\it no} explicit quasiperiodic forcing
and the dynamics is on SNAs.

These results are described in the following two sections
of this paper. In Section 2, we extend the general arguments
which establish the existence of SNAs in
Eqs.~(\ref{tanh}) and Eq.~(\ref{harpc}) so as to yield a large
variety of dynamical systems wherein the motion will
remain on SNAs.
In Section~3, we further generalize the dynamics so that
the forcing is no longer quasiperiodic but the resulting
dynamics continues to be on SNAs.  This is followed by a brief summary
and discussion in Section~4.

\section{Strange Nonchaotic dynamics}

By now there is a host of examples of strange nonchaotic
dynamics \cite{rmp} in driven systems. These include both flows
such as the driven Duffing oscillator \cite{duffing,duffing1} and
iterative maps such as the driven logistic map \cite{logistic,logistic1}.
Most studies to date have relied on numerical techniques to
establish the presence of SNAs, by explicitly computing
the Lyapunov exponents and by determining the fractal dimension.
SNAs are ``intermediate'' between strange chaotic attractors and
nonchaotic regular dynamics with regard to many dynamical
and structural properties, some of which can be determined
through study of power--spectra, correlation functions, and
related measures (see Ref.~\cite{rmp} for a recent review).
The most extensively studied cases all have a skew--product
dynamical structure, and thus these are systems in $n$+1
dimensions, the latter dimension pertaining to the dynamics
of the driving term.

In the present paper, we study the case of driven iterative
maps in 1+1 dimension for convenience.  Rigorous mathematical
results concerning SNAs are so far available only for the
simplest such examples. However, the arguments presented here
largely carry over to the case of flows as well, and can
be extended to higher dimensions, so this is not a restriction.

\subsection{Generalizing the original SNA system}
The following heuristic arguments, first put forward by
Grebogi \etl~\cite{gopy} for the system embodied in
Eq.~(\ref{tanh}) suggest that SNAs should occur
for appropriate values of $\alpha$.

The mapping $ x \to 2 \alpha \tanh x$ is 1--1
and contracting, taking the real line into the interval
$[-2\alpha, 2\alpha]$. Because $\omega$ is an irrational
number, the dynamics in $\phi$ [cf.  Eq.~(\ref{irra})] is
ergodic in the unit interval.  The attractor
of the dynamical system Eq.~(\ref{tanh}) therefore
must be contained in the strip  $[-2\alpha, 2\alpha]
\otimes [0,1]$.
A point $x_n$ with the corresponding $\phi_n =$ 1/4 will map to
($x_{n+1}=0, \phi_{n+1}= \omega + 1/4$), after which subsequent
iterates will all remain on the line ($x=0, \phi$), as will points
with $\phi_n = 3/4$. This  line
therefore forms an invariant subspace, but for large enough
$\alpha$, this subspace is transversally unstable.
Thus, it follows that the attractor has a dense
set of points on the line $x=0, \theta \in [0,1]$ (since $\omega$ is
irrational), but the entire line itself cannot be the attractor
for $\alpha >
1$, since the dynamics is unstable on that line.
There will, therefore at some $\alpha$,
be a ``blowout bifurcation'' \cite{yl,yl1} transition to
strange nonchaotic dynamics.

Generalizing Eq.~(\ref{tanh}) (keeping the skew--product
structure intact) as
\begin{eqnarray}
\label{tang}
x_{i+1} &=& \alpha f(x_i) g(\phi_i)
\end{eqnarray}
it is clear that the same arguments will carry over
so long as the following properties hold:

\begin{enumerate}
\item[(i)]
$f(x)$ is 1-1 and contracting, with $f(0)=0.$
\item[(ii)]
$f^{\prime} (0) \ne 0$, and concurrently,
\item[(iii)]
$g(\phi)= 0 $ for some $\phi = \phi_*$.
\end{enumerate}

Then, clearly, all points $x,\phi_*$ will map to $0,\{\omega
+\phi_*\}$, and the subsequent dynamics will be dense on the line
$x=0,\phi$. From condition (ii) above, this can be made
unstable locally for sufficiently large $\alpha$, and from (i),
since the map is contracting, the slope $\vert f^{\prime} \vert \le 1$
 almost everywhere, the
Lyapunov exponent can be made negative so as to give a SNA.

\subsection{A piecewise linear SNA}

A piecewise linear strange nonchaotic attractor can be simply
obtained by taking, in Eq.~(\ref{tang}),
\begin{eqnarray}
\label{lintanh}
f(x) &=& \alpha x ~~~~~~~~~~~~\vert x \vert \le 1/\alpha \nonumber \\
&=& \mbox{sign}(x) ~~~~~~\vert x \vert > 1/\alpha,\\
g(\phi) &=& \phi - 1/2.
\label{half}
\end{eqnarray}
It is easy to verify that there is indeed a blowout bifurcation
near $\alpha \approx 2.33$ (Fig.~1a), above which the attractor is
strange and nonchaotic: see Fig.~1b.

\subsection{Generalizing the Harper map}

In the Harper map, on the other hand, the argument for the
existence of strange nonchaotic dynamics
proceeds as follows \cite{ks}. In the strong--coupling limit,
$\alpha$ $\rightarrow$ $\infty$, Eq.~(\ref{harpc})
reduces to
\begin{eqnarray}
x_{i+1}=-[ 2\alpha \cos 2\pi\phi_i]^{-1}.
\end{eqnarray}
Clearly, for $\phi_{i}$ =1/4 or $\phi_{i}$ =3/4 this gives a
singularity in the neighbourhood of which the mapping
locally looks like a hyperbola.  Since the $\phi$
dynamics Eq.~(\ref{irra}) is ergodic in the
interval [0,1], the image of this singularity is dense:
on every $\phi$--fiber there will be a singularity, and thus the
dynamics is on a strange set. By continuity, even away
from the strong coupling limit but for $\alpha$ large enough,
this argument suggests that the dynamics can be strange. 
(In fact, for Eq.~(3), there are SNAs even at $\alpha = 1$.)
For a suitable choice of
function $f(x)$, the Lyapunov exponent may turn out to be negative;
in such a case, the attractor is strange {\it and} nonchaotic.
For the Harper map \cite{prss}, when $E = 0$ this is indeed the
situation.

Proceeding as above, one can generalize the Harper map
(with the $\phi$--dynamics unchanged) as
\begin{eqnarray}
\label{harpg}
x_{i+1} &=& [f(x_i)+ 2\alpha g(\phi_i)]^{-1}
\end{eqnarray}
where $f$ and $g$ are now arbitrary functions, the only additional
requirement being that $g(\phi)$ should have a zero in the interval
[0,1]. For suitable functions $f(x)$, the Lyapunov exponent can
indeed be made zero or negative, giving therefore, SNAs.

\subsection{SNAs of the Fibonacci chain}

The Harper map \cite{ks} derives from the Harper equation \cite{harper}
which is the discrete Schr\"odinger equation for a particle in a
quasiperiodic potential,
\beq
\psi_{n+1} + \psi_{n-1} + V(n) \psi_n = E \psi_n  \label{harpq},
\eqn
$\psi_n$ denoting the wave-function at lattice site $n$, the
potential being
$V(n) = 2\alpha \cos 2 \pi (n \omega + \phi_0)$.
The identification $\psi_{n-1}/\psi_n \equiv x_n$, connects
Eq.~(\ref{harpq}) and Eq.~(\ref{harpc}).
This system is known to have critical (or power--law localized) states
for $\alpha$ = 1,
when the classical system has critical SNAs \cite{nr1}. It is
also known that other forms of the potential $V(n)$ support
critically localized states \cite{kohmoto,kohmoto1,kohmoto2,pandit},
one example being the Fibonacci chain with
\beqr
\label{fibo}
V(n) &=& \alpha~~~~~~~~~~~~~ 0 \le \{n\omega\} \le \omega\\
&=& -\alpha~~~~~~~~~~~\omega < \{n\omega\} \le 1,
\end{eqnarray}
when all states are critical for any $\alpha$.

The classical map corresponding to this potential is
\begin{eqnarray}
\label{harpf}
x_{i+1} &=& -[x_i - E + V(i)]^{-1} \\
\phi_{i+1} &=& \{\omega+\phi_i\},
\end{eqnarray}
where $V(i)$ is given by Eq.~(\ref{fibo}), and there is an additional
overall phase $\phi_0$. It is a simple task to compute the Lyapunov
exponent for this mapping (Fig.~2a); if $E$ is an eigenvalue, then the
Lyapunov exponent is zero. Comparing Eq.~(\ref{harpg}) with
Eq.~(\ref{harpf}), the conditions on the functions
$f$ and $g$ are met, and therefore, the attractors of the above
system are SNAs; see Fig.~2b for an example.

\section{SNAs without quasiperiodic driving}

It is clear from the above constructions, that there are two main
ingredients in achieving SNAs in dynamical systems such as
Eq.~(\ref{tang}) or Eq.~(\ref{harpg}). Firstly, one needs some
mechanism for {\it local} instability while maintaining global
stability. The main purpose of the quasiperiodic driving, namely the
$\phi$ dynamics which is governed by Eq.~(\ref{irra}), is to make
the instabilities dense in $\phi$.

This suggests that a generalization of Eq.~(\ref{tang}) to
\begin{eqnarray}
\label{tangg}
x_{i+1} &=& \alpha f(x_i) g(\phi_i)\\
\phi_{i+1} &=& h(\phi_i),
\end{eqnarray}
or of Eq.~(\ref{harpg}) to
\begin{eqnarray}
x_{i+1} &=& -[f(x_i)+ \alpha g(\phi_i)]^{-1}\\
\phi_{i+1} &=& h(\phi_i),
\end{eqnarray}
where the function $h$ is not necessarily the rigid rotation,
but is otherwise such that the orbit $h^n(\phi_*)$ is dense
will still yield SNAs for sufficiently large $\alpha$.

To be mathematically more precise \cite{keller1}, we need that $h$ be 
a homeomorphism
with an invariant ergodic probability measure of full support,
and that the $h^{-1}$ orbit of $\phi_*$ be dense, in which
circumstances, the resulting attractor can be shown to be a SNA,
following the basic proof given by Keller \cite{keller} for the case
$f \equiv \tanh$, $g \equiv \cos$ and $h$ the irrational rigid rotation.
The basic property that is required is that the mapping $h(\phi)$ take the
interval $[0,2\pi]$ into some continuous subinterval (at least),
and that the orbit of a typical point should be dense in this subinterval.

It should be added that the function $h$ should have only nonpositive
Lyapunov exponents since the skew--product form for the dynamical system
is being retained. There are a number of possible choices for $h$
which are distinct from the quasiperiodic rotation, but which use
related maps to generate ergodic flows. This follows from the Weyl
theorem \cite{sinai} which states that if $\cal{H}$ is a polynomial
function of degree $r \ge 1$ with real coefficients $a_0, \ldots, a_r$,
at least one of which is an irrational number, then the map
\beq
\phi_n =\{ {\cal H}(n)\} \equiv \{a_0 + a_1 n + \ldots + a_r n^r\}
\eqn
is ergodic, and the sequence $\{\phi_n\}$ is uniformly distributed
in the interval [0,1]. If $\cal{H}$ is nonlinear, then it can be easily
verified that the sequence $\{\phi_n\}$ is not quasiperiodic, but
the Lyapunov exponent is zero since the map preserves distance.

An example of such a system is easy to devise. A simple choice is
to take ${\cal H}(n) = \omega n^2,\omega$ irrational, which gives
\begin{eqnarray}
\label{tangq}
x_{i+1} &=& \alpha f(x_i) g(\phi_i)\\
\phi_{i+1} &=& \{\phi_i + (2i+1)\omega \}.
\end{eqnarray}
The SNA which obtains for $f$ and $g$ given by
Eqs.~(\ref{lintanh}) and (\ref{half}),
for a suitably large value of $\alpha$ is shown in Fig.~3a.

Of course, it is also possible to use other mappings which generate
quasiperiodic motion to determine the $\phi$ dynamics. Many examples
of such maps are known, as for instance the diffeomorphisms
\beq
\phi \to \{\phi + \Omega + \epsilon a(\phi)\}
\eqn
for $\Omega$ a constant, $\epsilon$ sufficiently small and $a$
an arbitrary analytic
function.  The orbits are everywhere dense on the interval and
typically have irrational rotation number; indeed,
by Denjoy's theorem, any orientation preserving $C^2$ diffeomorphism
of the circle
is topologically equivalent to a rigid rotation \cite{arnold}. In the
specific case of the circle map with $a(\phi) \equiv \sin 2\pi\phi$
the parameter ranges wherein the map has irrational winding number
have been comprehensively described. 

Other possibilities for $h$ exist: there are examples of integrable
geodesic flows on compact manifolds which have positive topological
entropy but have no positive Lyapunov exponents \cite{keller1}, or
one can even take $h$ to be a SNA map such as Eq.~(\ref{tanh}) itself,
since that is known to provide an ergodic flow \cite{keller}. This
yields, for instance, the system
\begin{eqnarray}
x_{i+1} &=& 2 \alpha  \phi_i  \tanh x_i\\
\phi_{i+1} &=& 2 \beta \cos 2\pi \theta_i \tanh \phi_i\\
\theta_{i+1} &=& \{\omega+\theta_i\},
\end{eqnarray}
which has strange nonchaotic attractors for suitable values of
$\alpha$ and $\beta$, see Fig. 3b.

\section{Discussion and summary}

Although so far all known examples of systems with strange
nonchaotic dynamics have appeared to require quasiperiodic forcing,
the present work shows that this is, in fact, not necessary.
By suitably generalizing systems wherein SNAs are known to
exist \cite{ks,prss,keller,bo}, we have constructed
dynamical systems where the motion is on strange nonchaotic
attractors and there is no quasiperiodic driving.
This has particular relevance to experimental
observations of apparently nonchaotic attractors where
the dynamics does not have explicit quasiperiodic forcing.
One specific example is of a gas discharge plasma where the
light flux as a function of time (with current as the driving
parameter) appears to lie on a SNA. This has been
deduced by attractor reconstruction and extraction
of fractal dimension and Lyapunov exponents \cite{newexp}.
The other example \cite{newexp1} pertains to the effect of
noise which allegedly reproduces the effect of many--frequency
quasiperiodic driving.

In addition, we have also shown that strange nonchaotic dynamics
can be quite common.  Our approach has been heuristic:
the same deconstruction also provides a prescription for obtaining
a large variety of dynamical systems wherein SNAs must occur.
We have shown examples of two specific systems wherein the attractors
are piecewise--linear fractals. Detailed analyses of such examples
may prove to be simpler than the cases studied so far.

This procedure for creating SNAs is clearly not exhaustive
and there may be different
generalizations that will also yield strange nonchaotic dynamics.
In particular, we have chosen to stay within the ``skew--product''
class of mappings, although other possibilities \cite{brazil}
may be compatible with such dynamics as well.\\

\noindent
{\bf Acknowledgments}\\
We are indebted to Professor Gerhard Keller for advice and correspondence.
This research is supported by the Department of Science and Technology, India.

\newpage

\begin{figure}[htbp]
\epsfxsize=8.5cm
\centerline{\epsfbox{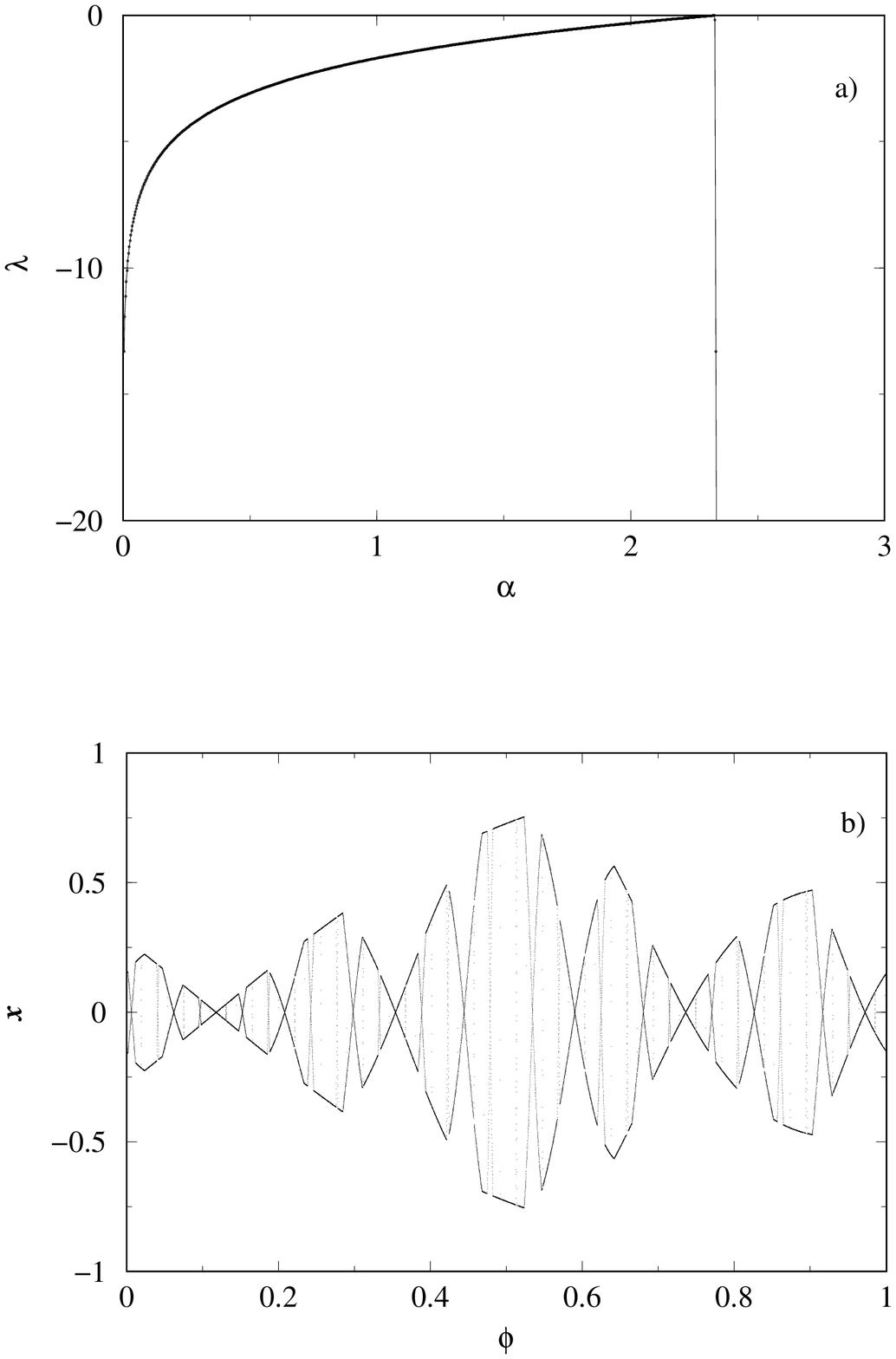}}
\caption{The piecewise linear system with $\omega = (\sqrt 5- 1)/2$.
Plot of (a) the Lyapunov exponent as a function of the parameter $\alpha$,
showing the blowout bifurcation at $\alpha \approx 2.33$, above which the
Lyapunov exponent $\to -\infty$, and (b) the strange non chaotic attractor
for $\alpha$=2.5}
\end{figure}

\newpage
\begin{figure}[htbp]
\epsfxsize=8.5cm
\centerline{\epsfbox{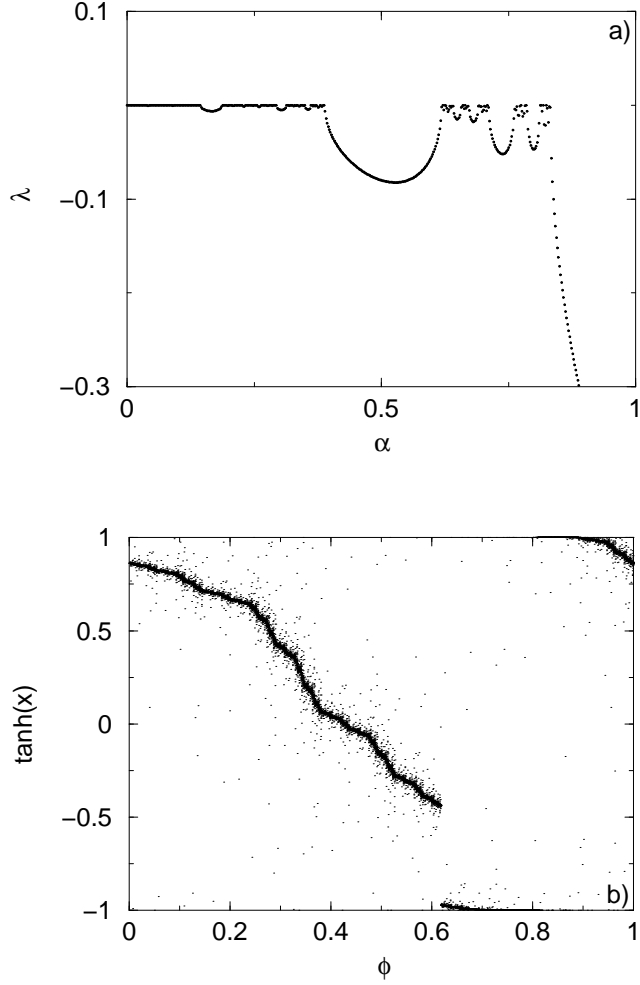}}
\caption{The Fibonacci chain, where $\omega = (\sqrt 5- 1)/2$. (a)
Variation of the Lyapunov exponent with forcing parameter $ \alpha$. 
Each zero of the Lyapunov exponent corresponds
to an eigenstate in the quantum system, and a critical SNA in the classical
system.  (b) A typical SNA in this system, for $E = 0$, $\alpha =0.8326745$.
Note that the variable plotted on the ordinate is $\tanh x$ rather than
$x$; this is merely for convenience.}

\end{figure}

\newpage
\begin{figure}[htbp]
\epsfxsize=8.5cm
\caption{(a) SNA in the system Eq.~(\ref{tangq}) where the driving is
no longer quasiperiodic. The parameters are $\alpha = 2.5, \omega $ is the
golden mean ratio. (b) SNA in the system given by Eqs.~(23--25),
where the map governing the $\phi$ dynamics itself has SNA dynamics.
The parameters $\alpha$ and $\beta$ are both 2.5.(this figure submitted in gif format)}
\end{figure}
\end{document}